\documentclass[final]{aipproc}

\layoutstyle{6x9}

\begin{document}

\newcommand\eps{{\cal E}}
\newcommand\fid{f_{\rm i}}
\newcommand\parn{\par\noindent}
\newcommand\pr{p_{\rm r}}
\newcommand\pt{p_{\rm t}}
\newcommand\psih{\Psi_0}
\newcommand\psis{\Psi_1}
\newcommand\psin{\Psi_{\rm N}}
\newcommand\psit{\Psi_{\rm T}}
\newcommand\psii{\Psi_{\rm i}}
\newcommand\psitil{\tilde\Psi}
\newcommand\psitilh{\tilde\psih}
\newcommand\psitils{\tilde\psis}
\newcommand\qi{Q_{\rm k}}
\newcommand\psitilt{\tilde\psit}
\newcommand\Qtil{\tilde Q}
\newcommand\ra{r_{\rm a}}
\newcommand\rai{r_{\rm ai}}
\newcommand\rc{r_{\rm c}}
\newcommand\roi{\rho _{\rm i}}
\newcommand\roqi{\varrho _{\rm i}}
\newcommand\rhotil{\tilde\rho}
\newcommand\rhotili{\tilde\rho_{\rm i}}
\newcommand\rhotils{\tilde\rhos}
\newcommand\rhotilh{\tilde\rhoh}
\newcommand\rotilq{\tilde\varrho}
\newcommand\sa{s_{\rm a}}
\newcommand\sigr{\sigma_{\rm r}}
\newcommand\sigt{\sigma_{\rm t}}

\title{On the global density slope--anisotropy inequality}

\classification{}
\keywords{}

\author{L. Ciotti}{
  address={Dept. of Astronomy, Univ. of Bologna,\\
via Ranzani 1, 40127 Bologna, Italy}
}

\author{L. Morganti}{
  address={Dept. of Astronomy, Univ. of Bologna,\\
via Ranzani 1, 40127 Bologna, Italy},
  ,altaddress={Max-Planck-Institut f\"ur Ex. Physik, Garching, Germany}
}

\begin{abstract}
  Starting from the central density slope--anisotropy theorem of An \&
  Evans \cite{ae06}, recent investigations have shown that the
  involved density slope--anisotropy inequality holds not only at the
  center, but at all radii (i.e. globally) in a very large class of
  spherical systems with positive phase--space distribution function.
  Here we present some additional analytical cases that further extend
  the validity of the global density slope--anisotropy inequality.
  These new results, several numerical evidences, and the absence of
  known counter--examples, lead us to conjecture that the global
  density slope--anisotropy inequality could actually be a universal
  property of spherical systems with positive distribution function.
\end{abstract}

\maketitle

\section{Introduction}

In the study of stellar systems based on the ``$\rho$--to--$f$''
approach (where $\rho$ is the material density and $f$ is the
associated phase--space distribution function, hereafter DF), $\rho$
is given, and specific assumptions on the internal dynamics of the
model are made (e.g. see \cite{ber}, \cite{bt08}).  In some special
cases inversion formulae exist and the DF can be obtained in integral
form or as series expansion (see, e.g., \cite{fri52}--\cite{cb05}).
Once the DF of the system is derived, a non--negativity check should
be performed, and in case of failure the model must be discarded as
unphysical, even if it provides a satisfactory description of data.
Indeed, a minimal but essential requirement to be met by the DF (of
each component) of a stellar dynamical model is positivity over the
accessible phase--space.  This requirement (also known as phase--space
consistency) is much weaker than the model stability, but it is
stronger than the fact that the Jeans equations have a physically
acceptable solution.  However, the difficulties inherent in the
operation of recovering analytically the DF prevent in general a
simple consistency analysis.

Fortunately, in special circumstances phase--space consistency can be
investigated without an explicit recovery of the DF.  For example,
analytical necessary and sufficient conditions for consistency of
spherically symmetric multi--component systems with Osipkov--Merritt
(hereafter OM) anisotropy (\cite{osi79}, \cite{mer85}) were derived in
\cite{cp92} (see also \cite{tre}) and applied in several
investigations (e.g., \cite{c96}--\cite{cmz}).  Moreover, in
\cite{cm09b} we derived analytical consistency criteria for the family
of spherically symmetric, multi--component generalized Cuddeford
\cite{cud} systems, which contains as very special cases constant
anisotropy and OM systems.

Another necessary condition for consistency of spherical systems is
given by the ``central cusp--anisotropy theorem'' by An \& Evans
\cite{ae06}, an inequality relating the values of the \textit{central}
logarithmic density slope $\gamma$ and of the anisotropy parameter
$\beta$ of \textit{any} consistent spherical system:

{\bf Theorem} In every consistent system with constant anisotropy
$\beta(r)=\beta$ necessarily
\begin{equation}
\label{ae}
\gamma(r)\equiv-\frac{d\ln\rho(r)}{d\ln r}\geq 2\beta
\quad\forall r,\quad{\rm where}\quad 
\beta(r)\equiv 1-\frac{\sigt^2(r)}{2\sigr^2(r)}.
\end{equation}
Moreover the same inequality holds asymptotically at the center of
every consistent spherical system with generic anisotropy profile.

In the following we call $\gamma(r)\geq 2\beta(r)$ $\forall r$ the
\textit{global} density slope--anisotropy inequality: therefore the An
\& Evans theorem states that constant anisotropy systems obey to the
global density slope-anisotropy inequality. However, constant
anisotropy systems are quite special, and so it was a surprise when we
found (\cite{cm09a}) that the necessary condition for model
consistency derived in \cite{cp92} for OM anisotropic systems can be
rewritten as the global density slope--anisotropy inequality. In other
words, the global inequality holds not only for constant anisotropy
systems, but also for each component of multi--component OM systems.
Prompted by this result, in \cite{cm09b} we introduced the family of
multi--component generalized Cuddeford systems, a class of models
containing as very special cases both the multi--component OM models
and the constant anisotropy systems.  We studied their phase--space
consistency, obtaining analytical necessary and sufficient conditions
for it, and we finally proved that the global density
slope--anisotropy inequality is again a necessary condition for model
consistency!

The results of \cite{cm09a} and \cite{cm09b}, here summarized,
revealed the unexpected generality of the global density
slope--anisotropy inequality.  In absence of counter--examples (see in
particular the Discussions in \cite{cm09b}) it is natural to ask
whether the global inequality is just a consequence of some special
characteristics of the DF of generalized Cuddeford systems, or it is
even more general, i.e. it is necessarily obeyed by all spherically
symmetric two--integrals systems with positive DF. Here we report on
two new interesting analytical cases of models, not belonging to the
generalized Cuddeford family, supporting the latter point of view. We
also present an alternative formulation of the global density--slope
anisotropy inequality.  Therefore, even if a proof of the general
validity of the global density slope--anisotropy inequality is still
missing, some relevant advance has been made, and we now have the
proof that entire new families of models do obey the global inequality
(see \cite{cm10} for a full discussion).

\section{The density slope--anisotropy inequality}
\subsection{Multi--component  Osipkov--Merritt systems}

The OM prescription to obtain radially anisotropic spherical systems
assumes that the associated DF depends on the energy and on the
angular momentum modulus of stellar orbits as
\begin{equation}
\label{fOM}
f(\eps,J)=f(Q),\quad Q=\eps-\frac{J^2}{2\ra^2},
\end{equation}
where $\ra$ is the so--called anisotropy radius (e.g. see
\cite{bt08}).  In the formula above $\eps=\psit-v^2/2$ is the relative
energy per unit mass, $\psit=-\Phi_{\rm T}$ is the relative (total)
potential, and $f(Q)=0$ for $Q\leq0$.  A multi--component OM system is
defined as the superposition of density components, each of them
characterized by a DF of the family~(\ref{fOM}), but in general with
different $\ra$.  Therefore, unless all the $\ra$ are identical, a
multi--component OM model is not an OM system.  It is easy to prove
that the radial dependence of the anisotropy parameter associated to
such models is
\begin{equation}
\label{betaOM}
\beta(r)=\frac{r^2}{r^2+\ra^2},
\end{equation}
i.e. systems are isotropic at the center and increasingly radially
anisotropic with radius.

Consistency criteria for multi--component OM models have been derived
in \cite{cp92}, while in \cite{cm09a} it was shown that a necessary
condition for phase--space consistency of each density component can
be rewritten as the global density slope-anisotropy inequality
\begin{equation}\label{cs}
\gamma(r)\geq 2\beta(r)\quad\forall r,
\end{equation}
i.e. not only constant anisotropy systems but also multi--component OM
models follow the global inequality.

\subsection{Multi--component generalized Cuddeford systems}

An interesting generalization of OM and constant anisotropy systems
was proposed by Cuddeford (\cite{cud}; see also \cite{ciotti}), and is
obtained by assuming
\begin{equation}
\label{f}
f(\eps,J)=J^{2\alpha}h(Q),
\end{equation}
where $\alpha>-1$ is a real number and $Q$ is defined as in
equation~(\ref{fOM}).  Therefore, both the OM models ($\alpha=0$), and
the constant anisotropy models ($\ra\to\infty$), belong to the
family~(\ref{f}).  In particular, it is easy to show that from
equation~(5)
\begin{equation}
\label{beta}
\beta(r)=\frac{r^2-\alpha\ra^2}{r^2+\ra^2}.
\end{equation}
Remarkably, also for these models a simple inversion formula links the
DF to the density profile (\cite{cud}). Such inversion
formula still holds for multi--component, generalized Cuddeford
systems, that we have introduced in \cite{cm09b}.  \textit{Each}
density component of a generalized Cuddeford model has a DF given by
the sum of an arbitrary number of Cuddeford DFs with arbitrary
positive weights $w_i$ and possibly different anisotropy radii $\rai$
(but same $h$ function and angular momentum exponent), i.e.
\begin{equation}
\label{sumCud}
f=J^{2\alpha}\sum_i w_i h(Q_i),\quad Q_i=\eps-\frac{J^2}{2\rai^2}.
\end{equation}
Of course, the orbital anisotropy distribution characteristic of
DF~(\ref{sumCud}) is \textit{not} a Cuddeford one, and quite general
anisotropy profiles can be obtained by specific choices of the weights
$w_i$, the anisotropy radii $\rai$, and the exponent $\alpha$.
However, near the center $\beta(r)\sim-\alpha$, and $\beta(r)\sim1$
for $r\to\infty$, independently of the specific values of $w_i$ and
$\rai$.

In \cite{cm09b}, we have found necessary and sufficient conditions for
the consistency of multi--component generalized Cuddeford systems.  At
variance with the simpler case of OM models, the new models admit a
\textit{family} of necessary conditions, that can be written as simple
inequalities involving repeated differentiations of the augmented
density expressed as a function of the total potential.
\textit{Surprisingly, we also showed that the first of the necessary
  conditions for phase--space consistency can be reformulated as the
  global density slope--anisotropy inequality (4)}, which therefore
holds at all radii for each density component of multi--component
generalized Cuddeford models.

\section{How general is the density slope--anisotropy inequality?}

The natural question posed by the analysis above is whether the global
density slope--anisotropy inequality is a peculiarity of
multi--component generalized Cuddeford models: after all, only models
in this (very large) family have been proved to obey the global
inequality.  We now continue our study by showing, by direct
computation, that two well-known anisotropic models, whose analytical
DF is available and not belonging to the generalized Cuddeford family,
indeed obey to the global density slope--anisotropy inequality.  A
full discussion of the following cases, and their place in a broader
context, will be presented in \cite{cm10}.

\subsection{The Dejonghe (1987) anisotropic Plummer model}

Dejonghe \cite{dej87}, by using the augmented density approach,
studied a family of (one--component) anisotropic Plummer models, with
normalized density--potential pair
\begin{equation}
\label{rhoDej}
\rho = \frac{3}{4\pi}\frac{\Psi^{5-q}}{(1+r^2)^{q/2}},\qquad
\Psi = \frac{1}{\sqrt{1+r^2}}.
\end{equation}
Both the radial trend of orbital anisotropy and the model DF were
recovered analytically:
\begin{equation}
\label{betaDej}
\beta(r)=\frac{q}{2}\frac{r^2}{1+r^2};\quad 
f=\eps^{7/2-q} g\left(\frac{J^2}{2\eps}\right),
\end{equation}
where $g$ belongs to the family of hypergeometric functions.  In
\cite{dej87} it is shown that the consistency requirement $f\geq0$
imposes the limitation $q\leq2$.  Well, a direct computation of the
logarithmic density slope of the Plummer model~(\ref{rhoDej}),
together with equation~(\ref{betaDej}), proves that these models obey
to the global density slope--anisotropy inequality when $q\leq2$.

\subsection{The Baes \& Dejonghe (2002) anisotropic Hernquist model}

Baes \& Dejonghe \cite{bd02} considered a family of one--component
anisotropic Hernquist models whose normalized density--potential pair
is
\begin{equation}
\label{rhoBaDej}
\rho =\frac{(1+r^2)^{2(\beta_0-\beta_\infty)}}{2\pi r^{2\beta_0}}\frac
{\Psi^{4-2\beta_0}}{(1-\Psi)^{1-2\beta_0}},
\qquad \Psi = \frac{1}{1+r},
\end{equation}
with $\beta_\infty\leq\beta_0$.
The corresponding anisotropy parameter and DF are
\begin{equation}
\label{betaBaDej}
\beta(r)=\frac{\beta_0+\beta_\infty r}{1+r};\quad 
f=\eps^{5/2-2\beta_\infty+\beta_0}J^{-2\beta_0} 
\sum_k\left(\frac{J^2}{2\eps}\right)^k g_k(\eps),
\end{equation}
so that $\beta_0$ and $\beta_\infty$ are the anisotropy values at the
center and at large radii of the system, respectively; note that in
this family of models the orbital anisotropy decreases moving away
from the center.  In equation~(\ref{betaBaDej}) $g_k$ are
hypergeometric functions and, in accordance with the ``cusp
slope--central anisotropy theorem'', the request of non--negativity
imposes $\beta_0\leq 1/2$ (see \cite{bd02}).  Note that, as in the
previous case, the DF is not of the generalized Cuddeford family.
Again a comparison of the logarithmic density slope of Hernquist
profile~(\ref{rhoBaDej}) with equation~(\ref{betaBaDej}) shows that,
when $\beta_\infty\leq\beta_0$ and $\beta_0\leq 1/2$ also these models
obey the global inequality (4)!

\subsection{Alternative formulation of the density slope--anisotropy
inequality}

While we refer the reader to \cite{cm10} for a full discussion of the
new results, and for how these find place in a more general context,
here we show that the density slope--anisotropy inequality can also be
expressed as a condition on the radial velocity dispersion.  In fact,
the relevant Jeans equation in spherical symmetry reads
\begin{equation}
\label{jeans}
\frac{d\rho\sigr^2}{dr}+\frac{2\beta\rho\sigr^2}{r}=\rho\frac{d\psit}{dr}
\end{equation}
(e.g., \cite{bt08}).
Introducing the logarithmic density slope and rearranging the terms,
one finds
\begin{equation}
\label{jeansGamma}
\gamma(r)-2\beta(r)=r\left(\frac{d\sigr^2}{dr}-\frac{d\psit}{dr}\right)\geq0
\end{equation}
as an equivalent, alternative formulation of the density
slope--anisotropy inequality.  Of course, the proof that a given
family of self--consistent models obeys inequality~(\ref{jeansGamma})
is not easier than the proof that would be obtained by working on
phase--space.

\section{Conclusions}

We have shown analytically that two more models, in addition to the
whole family of multi-component generalized Cuddeford systems, satisfy
the global density slope--anisotropy inequality as a necessary
condition for phase--space consistency.  This reinforces the
conjecture that the global slope--anisotropy relation (4) could be a
universal necessary condition for consistent spherical systems.  We
recall that additional evidences supporting such idea exist: for
example Michele Trenti kindly provided us with a large set of
numerically computed $f_{\nu}$ models \cite{bertre}, and all of them,
without exception, satisfy the inequality $\gamma(r)\geq2\beta(r)$ at
all radii.  Additional numerical findings are mentioned in
\cite{cm09b}.

\end{document}